\documentclass[a4paper,11pt]{article}
\usepackage{graphicx,amssymb,amstext,amsmath,mathabx,mathtools,esvect,ragged2e,floatrow}

\usepackage{floatflt}
\usepackage{subcaption}
\usepackage[dvipsnames]{xcolor}

\definecolor{cbl}{rgb}{0,0,1}                % bleu\

\newcommand{\bc}{\begin{center}}
\newcommand{\ec}{\end{center}}
\def\ba#1{\begin{array}{#1}\displaystyle}
\newcommand{\ea}{\end{array}}

\newcommand{\beq}{\begin{equation}}
\newcommand{\eeq}{\end{equation}}
\newcommand{\beqa}{\begin{eqnarray}}
\newcommand{\eeqa}{\end{eqnarray}}

\newcommand{\bi}{\begin{itemize}}
\newcommand{\ei}{\end{itemize}}
\newcommand{\p}{\partial}

\newcommand{\bra}{\langle}
\newcommand{\ket}{\rangle}

\newcommand{\TTb}{\mathrm{T}\overline{\mathrm{T}}}

\newcommand{\bal}{\boldsymbol{\alpha}}

\newcommand{\bel}{\boldsymbol{\beta}}

\usepackage{cite}

\definecolor{purple_nice}{rgb}{0.4,0.2,0.7}
\definecolor{fuel_blue}{RGB}{42,162,185}
\definecolor{YInMn_blue}{RGB}{46, 80, 144}
\definecolor{ultramarine}{RGB}{63, 0, 255}
\definecolor{KLEIN_blue}{rgb}{0, 0.18, 0.65}
\usepackage[linktocpage=true]{hyperref}
\hypersetup{
    colorlinks=true,
    linkcolor=YInMn_blue,
    citecolor=ultramarine,
    filecolor=fuel_blue,
    urlcolor=KLEIN_blue,
}
\begin{document}

\begin{titlepage}
\title{$\TTb$ Deformations and Form Factor Program}
\author{Olalla A. Castro-Alvaredo{\color{red} {$^\heartsuit$}}, Stefano Negro{\color{green} {$^\clubsuit$}} and Fabio Sailis{\color{blue} {$^\diamondsuit$}}}
\date{\small {{\color{red} {$^\heartsuit$}} \color{blue} {$^\diamondsuit$}} Department of Mathematics, City St George's, University of London,\\ 10 Northampton Square EC1V 0HB, UK\\
\medskip
{\color{green} {$^\clubsuit$}}  Department of Mathematics, University of York,
Heslington, York YO10 5DD, UK\\
\medskip
}
\maketitle
\begin{abstract}
In this proceeding contribution, we review a recently proposed method to compute the minimal form factors (MFFs) of diagonal integrable field theories perturbed by irrelevant fields of the $\TTb$ family. Our construction generalizes standard form factor techniques to deal with the deformed two-body scattering amplitudes, which are typical in this setting. The results are minimal form factors which are the product of the undeformed solution and a new function. This function can be fixed by requiring constant asymptotics for large rapidities, smoothness in the limit when the perturbation parameters go to zero, and agreement with standard MFF formulae for particular choices of the perturbation couplings. We observe that, for a certain range of parameters, the new MFF develops a pole at $\theta=0$. By considering several UV-complete theories, we argue that such poles can emerge naturally from the MFF integral representation and suggest how they may be eliminated.  
\end{abstract}


\vfill

\noindent 
{\Large {\color{red} {$^\heartsuit$}}}o.castro-alvaredo@city.ac.uk\\
{\Large {\color{green} {$^\clubsuit$}}}stefano.negro@york.ac.uk\\
{\Large {\color{blue} {$^\diamondsuit$}}}fabio.sailis@city.ac.uk\\

\hfill \today
\end{titlepage}

\section{Introduction}
Deformations of 2D quantum field theory (QFT) via irrelevant operators of the $\TTb$ family \cite{Smirnov:2016lqw, Cavaglia:2016oda, Dubovsky:2012wk, Caselle:2013dra} have been extensively studied in the literature. $\TTb$ perturbations preserve integrability and give rise to models with interesting new properties. For this reason, there are many works on $\TTb$ perturbations and their generalizations, particularly in the context of 1+1D integrable quantum field theory (IQFT) \cite{Smirnov:2016lqw, Cavaglia:2016oda, Conti:2018jho, Conti:2018tca, Conti:2019dxg, Dubovsky:2023lza} and via the thermodynamic Bethe ansatz (TBA) approach \cite{Hernandez-Chifflet:2019sua, Camilo:2021gro, Cordova:2021fnr, LeClair:2021opx, LeClair_2021, Ahn:2022pia, Dubovsky:2012wk, Caselle:2013dra,RMO}. There are many other viewpoints which we will not discuss here due to length restrictions. Our focus will be only on IQFTs where powerful techniques are at our disposal \cite{Zamolodchikov:1989hfa, Smirnov:2016lqw,Negro:2016yuu}. In particular, it is well known that $\TTb$-like perturbations modify the exact two-body scattering matrix by a multiplicative (CDD) factor \cite{Castillejo:1955ed}. Throughout this proceeding, we will consider theories with a single particle spectrum. This means that the scattering matrix and scattering phase carry no particle indices.  Then,  the deformed $S$-matrix is
\begin{equation}
    S_{\boldsymbol{\alpha}}(\theta) = \Phi_{\boldsymbol{\alpha}}(\theta) S(\theta)\;,
    \label{Smatrix}
\end{equation}
where $\boldsymbol{\alpha} = (\alpha_s)_{s\in\mathcal{S}\subset\mathbb{N}}$ is a, possibly infinite, vector, $S(\theta)$ is the two-particle scattering matrix for a theory with a single particle, and
\begin{equation}
    \Phi_{\boldsymbol{\alpha}}(\theta) = \exp\left[-i\sum_{s\in \mathcal{S}} \alpha_s m^{2s} \sinh(s\theta)\right]\;.
    \label{sum}
\end{equation}
Here $m$ is a fundamental mass scale such that the combination $\alpha_s m^{2s}$ is dimensionless. We will take $m=1$ for simplicity. $\mathcal{S}$ is a set of spin values, typically those of local conserved charges. Notice that $\mathcal{S}$ has to be a subset of the odd integers, otherwise the CDD factor does not satisfy the crossing equation $\Phi_{\boldsymbol{\alpha}}(i\pi-\theta) = \Phi_{\boldsymbol{\alpha}}(\theta)$. Since $\Phi_{\boldsymbol{\alpha}}(\theta)$ is a CDD factor, the theory described by this new $S$-matrix is still integrable and has the same particle spectrum as the original model.   

In our papers \cite{longpaper, ourTTb3,  Castro-Alvaredo2024compl,ourboundary} we worked on the problem of computing matrix elements of ``local" operators via the form factor bootstrap program \cite{KW, smirnov1992book}. The $n$-particle form factor of an operator $\mathcal{O}$ between the ground state $|0\ket$ and an $n$-particle in-state of paricles of the same species and distinct rapidities $\theta_1 \ldots \theta_n$ is usually written as,
\beq 
F_n^{\mathcal{O}}(\theta_1\dots,\theta_n):=\bra 0| \mathcal{O}(0)|\theta_1, \dots \theta_n|0\ket\,.
\label{introFF}
\eeq 
The form factor program is a method for computing these objects as solutions to a set of consistency equations. In our work \cite{longpaper, ourTTb3, Castro-Alvaredo2024compl,ourboundary} we pursued this program and found closed formulae for the form factors of a large class of fields and theories while also leaving several open questions. One of the most important open questions relates to the building block of all form factor solutions, a function known as the minimal form factor (MFF). In this proceeding contribution, we review the construction of this MFF, following mainly \cite{Sailis:2025kvn}. In the conclusion and discussion section, we present some new insights into this solution.
\section{The Minimal Form Factor}
The minimal form factor is a cornerstone of the form factor program and typically the first non-trivial function one must compute in pursuit of higher particle solutions. It is a minimal solution (ie. without poles in the physical strip ${\rm Im}(\theta)\in [0,\pi]$) of the form factor equation for two particles ($n=2$). In our notation, this would be $F_n^{\mathcal{O}}(\theta_1,\theta_2)$, however, in relativistic models and for spinless fields, the form factors can only depend on rapidity differences. Therefore, the two-particle form factor depends on a single rapidity variable. For a perturbed theory, it will also depend on the parameters $\bal$, so we will denote it by $F_{\rm min}(\theta;\bal)$. From the form factor equations that can be found, for example, in \cite{longpaper}, we have that
\beq 
F_{\rm min}(\theta;\bal)=S_{\bal}(\theta)F_{\rm min}(-\theta;\bal)=F_{\rm min}(2\pi i-\theta;\bal)\;.
\label{RHproblem}
\eeq 
The MFF is then an entire solution to these equations.
The same equations with $S_{\bal}(\theta)$ replaced by $S(\theta)$ are satisfied by the MFF of the undeformed theory,  $F_{\rm min}(\theta)$:
\beq 
F_{\rm min}(\theta)=S(\theta)F_{\rm min}(-\theta)=F_{\rm min}(2\pi i-\theta)\;.
\label{FFE}
\eeq
This means that the MFFs $F_{\rm min}(\theta;\bal)$ and $F_{\rm min}(\theta)$ are proportional to each other through a function $\mathcal{D}_{\bal}(\theta)$ which satisfies
\beq
F_{\rm min}(\theta;\bal) =  \mathcal{D}_{\bal}(\theta) F_{\rm min}(\theta)\quad \Longrightarrow \quad \mathcal{D}_{\bal}(\theta)=\Phi_{\bal}(\theta) \mathcal{D}_{\bal}(-\theta)=\mathcal{D}_{\bal}(2\pi i-\theta)\;.
\label{DMM}
\eeq 
Note that each of the first equalities in (\ref{RHproblem})-(\ref{DMM}) defines a Riemann-Hilbert problem for the corresponding functions, with $S_{\bal}(\theta)$, $S(\theta)$ and $\Phi_{\bal}(\theta)$ playing the role of ``jump functions" \cite{gakhov2014boundary}. This can be used to explicitly write the solution as an integral representation involving the logarithm of the jump function, as we will discuss in the next section. It has been shown that the function $\mathcal{D}_{\bal}(\theta)$,  has the factorised structure \cite{Castro-Alvaredo2024compl,longpaper}
\beq 
\mathcal{D}_{\bal}(\theta):= \varphi_{\bal}(\theta)C_{\bel}(\theta)\;,
\label{minimal}
\eeq
with
\beq 
\varphi_{\bal}(\theta)=\exp\left[{\frac{
\theta-i\pi}{2\pi} \sum_{s\in \mathcal{S}} \alpha_s \sinh(s\theta)}\right]\;, \qquad 
C_{\bel}(\theta):=\exp\left[\sum_{n\in \mathbb{Z}^+} \beta_n  \cosh(n\theta) \right]\;.
\label{betas}
\eeq 
Here the parameters $\alpha_s$ in (\ref{betas}) are those defining the CDD factor $\Phi_{\bal}(\theta)$ \eqref{sum}, while the parameters $\beta_n$ are, a priori, free and independent of the scattering phase. The existence of this very large freedom in the choice of the MFF is an ambiguity that was already highlighted in the original works \cite{ourTTb3,longpaper,Castro-Alvaredo2024compl}, where we chose $\beta_n=0$ for all $n$. In a subsequent paper \cite{MMF}, we showed that the sinh-Gordon theory can also be seen as the Ising model perturbed by infinitely many irrelevant perturbations with finely-tuned couplings $\alpha_s$ (see Eq.~(\ref{alphas}) in Section~\ref{conc}). Then, its MFF, which is usually obtained from the integral representations we discuss in Section \ref{intrep} can be rewritten in the form \eqref{minimal} where the ``free" parameters $\beta_n$ are now fixed in terms of $\alpha_s$. This generalizes easily to other IQFTs. In these examples, it is clearly seen that the role of the function $C_{\bel}(\theta)$ is to tame the unphysical asymptotic properties of the function $\varphi_{\bal}(\theta)$ whilst ensuring analyticity, producing a well-defined MFF that is compatible with UV-completeness of the theory. We will show in Section~\ref{intrep} that the function $C_{\bel}(\theta)$ can also be fully fixed for theories perturbed by a finite number of irrelevant perturbations. 

\section{Integral Representations}
\label{intrep}
Given a massive, UV-complete IQFT with $S$-matrix $S(\theta)$ it was shown in \cite{KarowskiWeisz:1978vz} that the MFF is uniquely fixed by the combined requirements of analyticity in the physical strip and (at most) exponential growth for $\theta$ large.
The integral representation takes a particularly simple form in terms of the Fourier transform of the $S$-matrix phase:
\beq
i\delta(\theta):=\log{S(\theta)}= \int_0^{\infty}\frac{d\,t}{t} g(t) \sinh{\frac{t \theta}{i \pi}}
\label{Smatrixfourier}
\eeq
then
\beq
\rho(\theta):=\log{F_{\min}(\theta)}= \int_0^{\infty}\frac{d\,t}{t} \frac{g(t) }{\sinh{t}} \sin^2\frac{t(i\pi-\theta)}{2\pi}\,.
\label{KWFourierMFF}
\eeq
The challenge for $\TTb$-perturbed theories lies precisely in finding convergent representations of this type. This is due to the distinct properties of the CDD-factor (\ref{sum}). The difficulties can be best appreciated and fixed by starting with a different, yet equivalent, version of the integral representation, given explicitly in terms of $\log S(t)$. This kind of representation appeared in \cite{niedermaier1995freefieldrealizationform}:
\beq
\rho(\theta) =   \frac{1}{i \pi} \cosh^2{\frac{\theta}{2}} \int_0^{\infty} dt \frac{ \tanh\frac{t}{2} \ln{S(t)}}{ \cosh{t} - \cosh{\theta} }   \,, \qquad \textrm{for}\; 0 \leq {\rm Im}(\theta) \leq 2 \pi \,.
\label{NiedermeierMFF}
\eeq
This is equivalent to \eqref{KWFourierMFF} once we invert the Fourier transform. Consider now $\Phi_{\bal}(\theta)$ as defined in (\ref{sum}). We have 
\beqa
\log{\mathcal{D}_{\bal}(\theta)}= \frac{1 }{i \pi} \cosh^2{\frac{\theta}{2}} \int_0^{\infty} dt \frac{ \tanh \frac{t}{2} \ln{\Phi_{\bal}(t)}}{ \cosh{t} - \cosh{\theta} }= -\frac{1}{ \pi} \cosh^2{\frac{\theta}{2}} \sum_{s \in \mathcal{S}} \int_0^{\infty} dt \frac{\alpha_s \tanh\frac{t}{2}\sinh{st}}{ \cosh{t} - \cosh{\theta}}\,.
\label{62}
\eeqa
This integral is not convergent because of the asymptotics of the $\sinh(s t)$ function. We can easily see this by computing the integral as follows. 
For $| \textrm{Re}(\theta)|<\textrm{Re}(t)$  the integrand can be expanded as a long-wave expansion in $\theta$  \cite{Zamolodchikov:1991MasslessFlows, niedermaier1995freefieldrealizationform} giving
\beq
\frac{\cosh^2{\frac{\theta}{2}}\tanh{\frac{t}{2}}}{\cosh{t}-\cosh{\theta}}= \sum_{m = 1}^{\infty} \left[ \cosh(m \theta)+(-1)^{m+1} \right] e^{-m t}\,.
\label{longwaveexp}
\eeq
Consider next the case of a single perturbation of spin $s$.  We get then 
\beq
\log{\mathcal{D}_{\alpha_s}(\theta)}= -\frac{\alpha_s}{ \pi} \sum_{m= 1}^{\infty} \left[ \cosh{m \theta}+(-1)^{m+1} \right] \int_0^{\infty} dt  e^{-m t} \sinh{st}\,.
\eeq
We are left with the computation of a simple Laplace transform
$
 \int_0^{\infty} dt  e^{-m t} \sinh{st}= \frac{s}{m^2-s^2}
 $, 
which is only well defined away from the simple pole at $m>s$. This means that we need to regularize the infinite sum in $m$ of \eqref{longwaveexp}
by subtracting the pole. To this end, we modify the Laplace transform to 
\beq
\frac{s}{m^2-s^2} -\frac{m}{m^2-s^2}= - \frac{1}{m+s}\,,
\label{66}
\eeq
which is equivalent to replacing $\sinh(s t)$ by $-e^{-s t}$. This amounts to the regularisation prescription
\beq
\log{\mathcal{D}_{\bal}(\theta)}=  \frac{1}{ i \pi}  \cosh^2{\frac{\theta}{2}} \int_0^{\infty} dt \frac{\tanh\frac{t}{2} }{ \cosh{t} - \cosh{\theta} } \left( \ln{\Phi_{\bal}(t)}-  \p_t \ln{\hat{\Phi}_{\bal}(t)} \right)\;,
\label{niceformula}
\eeq
where 
\beq 
\ln \hat{\Phi}_{\bal}(\theta)=-i \sum_{s\in 2 \mathbb{Z}^{+} -1} \frac{\alpha_s}{s} \sinh(s\theta)\,.
\label{phihat}
\eeq 
Taking $s=2n-1$ (odd spin),  the regularised version reads
\beq
\log{\mathcal{D}_{\alpha_{2n-1}}(\theta)}= \frac{ \alpha_{2n-1}}{ \pi} \sum_{m = 1}^{\infty} \frac{\cosh(m \theta)+(-1)^{m+1}}{m+2n-1} \,.
\label{79}
\eeq
The sum above can be evaluated exactly (see \cite{Sailis:2025kvn} for the details). 
Putting everything together: 
\beqa 
\log\mathcal{D}_{\alpha_{2n-1}}(\theta)&=&\frac{\theta-i\pi}{2\pi} \alpha_{2n-1} \sinh((2n-1)\theta)+\frac{\alpha_{2n-1}}{\pi}\left(- \log 2 + c_{2n-1}\right)\\
&& -\frac{\alpha_{2n-1}}{\pi}\sum_{m=1}^{2n-2} \frac{\cosh(m\theta)}{2n-1-m}-\frac{\alpha_{2n-1}}{2 \pi} \cosh((2n-1)\theta) \log\left(-4\sinh^2\frac{\theta}{2}\right)\,,\nonumber
\eeqa 
where $c_{2n-1}:=\sum_{m=1}^{2n-2}\frac{(-1)^{m+1}}{m}$.
Generalizing to any number of $\TTb$ perturbations and employing the standard normalization at $\theta=i\pi$ we can write a new general formula without free parameters for the function $C_{\bel}(\theta)$ defined in (\ref{betas}):
\beq 
C_{\bel}(\theta)=C_{\bel}(i\pi) \prod_{s \in 2\mathbb{Z}^{+}-1}\left[-\frac{1}{2}\,e^{c_s-\sum\limits_{m=1}^{s-1}\frac{\cosh(m\theta)}{s-m}} \left(2i \sinh{\frac{\theta}{2}} \right)^{-\cosh{(s\theta)}}\right]^{\frac{\alpha_s}{\pi}}\,.
\label{newC}
\eeq 
and $\varphi_{\bal}(\theta)$ is the same function as in \eqref{betas}. It is easy to show that, for $|\theta|$ large, this function tends to $e^{-\frac{\alpha_{\hat{s}}|\theta|}{2\pi}e^{\hat{s}|\theta|}}$ with $\hat{s}$ the largest spin involved in the product. Combining this scaling with that of the function $\varphi_{\bal}(\theta)$ in (\ref{betas}) we see that the function $
\mathcal{D}_{\bal}(\theta)$ tends to a constant (independent of $\theta$) for large $\theta$. This means that the scaling of the MFF for large rapidity is the same as for the unperturbed theory.
This ``tamed" asymptotics is a nice feature of our solution, which stands in contrast to the very rapid growth (or suppression, depending on the sign of $\alpha_{\hat{s}}$) that we found in our earlier work \cite{longpaper,Castro-Alvaredo2024compl}, where we had instead chosen $C_{\bel}(\theta)=1$.
\section{Example: The Ising Model Perturbed by $\TTb$}
\begin{figure}
\floatbox[{\capbeside\thisfloatsetup{capbesideposition={right,top},capbesidewidth=6cm}}]{figure}[\FBwidth]
{\caption{The modulo square of the MFF of the Ising field theory perturbed by $\TTb$ for different values of the coupling $\alpha$. The black line correspond to $\alpha=0$. Other lines correspond to values $\pi, \pi/2, \pm 2\pi$ and $3\pi$. It is important to note that there is a change at $\alpha=\pi$. For $\alpha>\pi$ the MFF diverges at $\theta=0$ whilst for $\alpha\leq \pi$ it has a zero, just like for the unperturbed model.}\label{fmin}}
{\includegraphics[width=8cm]{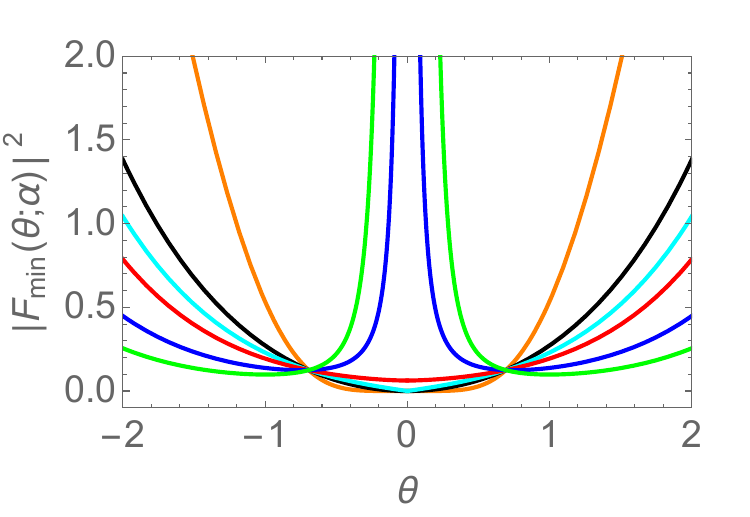}}
\end{figure}

In this section we analyze the properties of the new MFF by considering the simplest example of its construction. We consider the Ising field theory, which has two-body scattering matrix $S(\theta)=-1$ and MFF given by \cite{YZam}
\beq 
F_{\rm min}(\theta)=-i\sinh\frac{\theta}{2}\,.
\label{ising}
\eeq 
Therefore, if we perturbed the Ising field theory by a single $\TTb$ perturbation, corresponding to spin $s=1$ and coupling $\alpha:=\alpha_1$, the formula (\ref{newC}) from the previous section, together with (\ref{betas}) give us the new MFF as
\beq 
F_{\rm min}(\theta;\alpha)= (-2)^{-1-\frac{\alpha}{\pi}}e^{\frac{\theta-i\pi}{2\pi} \alpha \sinh\theta}\left(2i\sinh\frac{\theta}{2}\right)^{1-\frac{\alpha}{\pi}\cosh\theta}\,.  
\label{Ifor}
\eeq 
In the context of the form factor program, the MFF is a building block for all form factors and enters the computation of correlation functions through its modulo squared. This is the function shown in Figure~\ref{fmin}. As we see in the figure, and also discuss in the caption, the MFF has either a zero at $\theta=0$, as for the unperturbed Ising model, or a pole, when the power of $\sinh\frac{\theta}{2}$ becomes negative. This happens for $\alpha>
\pi$ in this case. Therefore, there is a critical value of $\alpha$ for which there is a qualitative change in the behaviour of the MFF. Since the pole at $\theta=0$ is nonphysical, it is interesting to explore why it is present in some cases. We present some additional discussion in our final section. 
\section{Conclusion and Discussion}
\label{conc}
We have discussed the construction of consistent MFFs for 1+1D IQFTs, summarizing mainly the results of our recent paper \cite{Sailis:2025kvn}. Our main result is an MFF that has good asymptotic and reduction properties. By this we mean that, growth with the rapidity variable $\theta$ is exponential at most and dictated by the unperturbed MFF. ``Reduction" is meant in two senses. First, that the perturbed MFF reduces to the unperturbed one when all $\alpha_s=0$. Second, which we have not discussed here, that it reduces to a new functional representation of known MFFs of IQFTs when the number of $\alpha_s$ parameters is infinite and they take precise, theory-dependent values. 

In connection to the above, we would like to conclude by adding a short discussion here, which puts the presence of a zero/pole at $\theta=0$ in our MFF in the more general context of IQFTs. This is a new observation, which has not appeared elsewhere.
The presence of a zero at $\theta=0$ is common to many IQFTs with a fermionic $S$-matrix. We find this also in the sinh-Gordon model, for instance \cite{Fring_1993,MMF}, since the ``Ising factor" $\sinh\frac{\theta}{2}$ is also present as a prefactor,  even if the full form factor is much more complicated than (\ref{Ifor}). As discussed in \cite{MMF}, the $S$-matrix of the sinh-Gordon model admits a formal representation of the type (\ref{Smatrix}), with
\beq 
\alpha_{s}= \frac{4 i^{s+1}}{s} \,{\cos \frac{s b \pi}{2}}\,,\quad \mathrm{with} \quad s\in 2\mathbb{Z}^+ -1\,,
\label{alphas}
\eeq 
where $b$ is a coupling constant, a characteristic of the model. In this representation, the unperturbed theory is the Ising model, with its MFF given above (\ref{ising}). This means that for $\theta=0$ the sinh-Gordon model MFF has a zero of order given by
\beq 
1-\sum_{s\in 2\mathbb{Z}^+ -1}\frac{4 i^{s+1}}{\pi s} \,{\cos \frac{s b \pi}{2}}=\left\{ \begin{array}{ccc}
2&\mathrm{for}& b\in\mathbb{R}\quad \mathrm{and}\quad  0<|b|<1\\
1&\mathrm{for}& b\in\mathbb{R}\quad \mathrm{and}\quad \qquad |b|= 1\\
0&\mathrm{for}& b\in\mathbb{R}\quad \mathrm{and}\quad \qquad |b|> 1\\
f(\theta_0)&\mathrm{for}& \, \, \, b\in\mathbb{C}\quad \mathrm{and}\quad \qquad b=\frac{2i\theta_0}{\pi}\\
\end{array}\right.
\label{power}
\eeq 
These values reveal some interesting properties of standard IQFTs:
\begin{itemize}
\item The case $0<|b|<1$ is the natural range of coupling values for the sinh-Gordon model. There is a zero at $\theta=0$, which is well known from the standard MFF representation \cite{Fring_1993}. 
\item For $b=\pm 1$ the sinh-Gordon $S$-matrix reduces to $1$, that is, it is the free-boson limit. In this limit the MFF of the ``unperturbed" theory is no longer that of Ising but that of a free boson, which is also $1$. In other words, for $b=\pm 1$ the power one of $\sinh\frac{\theta}{2}$ is absent, and we have neither a zero nor a pole at $\theta=0$. 

\item For $|b|>1$ we exit the allowed values in the sinh-Gordon theory and enter a regime which contains in particular the $S$-matrix of the Lee-Yang model \cite{LYCardy}. The latter corresponds to the choice $|b|=5/3$ and is also a fermionic theory, so the power (\ref{power}) is present in the MFF, which is 0 at $\theta=0$. The limit $\theta\rightarrow 0$ is, in fact, a little tricky, giving the indeterminate $0^0$. The known MFF of the Lee-Yang model \cite{LYFFs} contains a modification that avoids this singularity. In \cite{LYFFs} a MFF was proposed which contains an additional multiplicative factor $\frac{\cosh\theta-1}{\cosh\theta-\cos\frac{\pi}{3}}$. This has the double effect of canceling the singularity at $\theta=0$ and of introducing a bound state pole, which is necessary in this theory. 
\item Finally, for imaginary $b$, we recover the roaming trajectories model, a theory whose RG-flow is known to visit the vicinity of infinitely many critical points corresponding to the unitary minimal models of CFT \cite{roaming}. In this case the cosine in the sum above becomes $\cosh(s\theta_0)$, giving a divergent sum. We can however, still make sense of (\ref{power}) if we appeal to a property of the model, namely symmetry under  $\theta_0 \mapsto -\theta_0$. Based on this property, we can argue that we might replace $\cosh(s\theta_0) \mapsto e^{-s|\theta_0|}$. A similar kind of argument was also employed in the form factor calculations presented in \cite{Horvath:2016nbm} and termed CPT symmetrization. This step allows us to obtain a sum which is convergent and gives power
\beq 
f(\theta_0)=\left\{ \begin{array}{ccl}
2&\mathrm{for}& |\theta_0|\,\, \mathrm{finite} \\
1& \mathrm{for}& |\theta_0|\rightarrow \infty\\
\end{array}\right.
\label{power}
\eeq 
Therefore the roaming trajectories model has a MFF with a zero at the origin. The order of the zero is just 1 in the $|\theta_0|\rightarrow \infty$ limit, which corresponds to the Ising field theory.
\end{itemize}
Equation (\ref{power}) shows that, for UV-complete fermionic IQFTs, the allowed $S$-matrices must satisfy strict constraints leading to positive powers in the absence of bound state poles and zero or negative powers in the presence of bound states. It would be interesting to investigate whether one could start from these constraints and work backwards to systematically classify all UV complete IQFTs, similar to what was done via thermodynamic Bethe ansatz in \cite{Ahn:2022pia}.\\

\noindent {\justifying {\bf Acknowledgments:} Olalla A. Castro-Alvaredo and Fabio Sailis are grateful to the organizers of the {\it XXIX International Conference on Integrable Systems and Quantum Symmetries}, held in Prague (Czech Republic), July 7-11 (2025), for giving us the opportunity to present this work and to contribute to these Proceedings. Fabio Sailis is grateful to the School of Science and Technology of City St George's, University of London, for a Ph.D. Studentship.}

\bibliography{bibliography}
\end{document}